\documentclass[]{jkas} 

\def\year{2026} 
\def\volume{56} 
\def\issue{1} 
\def\beginpage{1} 
\newcommand{\kms}{\rm km~s^{-1}}

\newcommand{\Msun}{\rm M_{\odot}}
\usepackage{float}

\title{Weak Lensing Spectrotomography: A1767 and A2065}

\author[1,2]{Jubee Sohn}{0000-0002-9254-144X}
\author[3]{Ian P. Dell'Antonio}{0000-0003-0751-7312}
\author[4]{Margaret J. Geller}{0000-0002-9146-4876}

\affil[1]{Department of Physics and Astronomy, Seoul National University, Gwanak-gu, Seoul 08826, Republic of Korea}
\affil[2]{SNU Astronomy Research Center, Seoul National University, Seoul 08826, Republic of Korea}
\affil[3]{Physics Department, Brown University, 182 Hope Street, Providence, RI 02912}
\affil[4]{Smithsonian Astrophysical Observatory, 60 Garden Street, Cambridge, MA 02138, U.S.A}

\def\corrauthor{J. Sohn, \email{jubee.sohn@snu.ac.kr}}

\def\runningauthor{Sohn, Dell'Antonio \& Geller}

\def\runningtitle{Weak Lensing Spectrotomography}

\def\keywords{galaxies: clusters: general, galaxies: distances and redshifts, galaxies: haloes, surveys}

\def\abstracttext{We describe spectroscopic tomographic weak lensing measurements (spectrotomography) for two rich clusters of galaxies, A1767 and A2065, based on extensive spectroscopy and Subaru/Hyper Suprime-Cam (HSC) imaging. These detections represent the first use of spectrotomography based on archival Subaru/HSC imaging. The measurements depend only on galaxies with spectroscopic redshifts reported here. The approach cleanly separates cluster members from the background and suppresses systematics that may be introduced by the use of photometric background redshifts. We detect the tomographic shear signals at $3.1\sigma$ (A1767) and $3.5\sigma$ (A2065). The shear signal amplitudes are consistent with the cluster dynamical (caustic) masses and they scale appropriately with source redshift. However, comparison with the first spectrophotometric detection of A2029 based on DECam imaging reveals some subtle potential systematic issues in deriving the shear signal for the relatively bright background galaxies used in the analysis. These issues may be important for understanding future more extensive applications of spectrotomography based on further Subaru imaging, as well as Euclid and LSST data. The total of three spectrophotometric detections (A1767, A2029, and A2065) sets the stage for broader application of the technique for unbiased cluster weak lensing mass determinations and potentially for a geometric cosmological test that is independent of other methods.}

\begin{document}
\jkashead 


\section{INTRODUCTION\label{sec:intro}}

Large format cameras on large telescopes in excellent sites have enabled weak lensing studies of clusters at redshift $z \lesssim 0.12$ (e.g., \citealp{Fu22, Fu24, HyeongHan24, HyeongHan25, Kim26}). These observations have already provided insights into cluster masses and concentrations (e.g., \citealp{Herbonnet22}), filaments that link neighboring clusters to one another (e.g., \citealp{HyeongHan24}), and to the relationships among the cluster mass distribution, the cluster BCG, and the galaxy population (e.g., \citealp{Montes21, JimenezTeja24}). These investigations set the stage for even more ambitious observations and analyses based on sky surveys with, for example, Euclid (e.g., \citealp{Ingoglia24}) and the Vera C. Rubin Observatory \citep{lsst19}.

The power of weak lensing for exploring fundamental astrophysical issues has some limitations. These limitations include lack of precise determination of the redshift distribution of background sources and tentative contamination by cluster members in the lensing sample. To mitigate these limitations, a variety of techniques for separating members and background sources based on colors and photometric redshift have been applied. However, large uncertainties in the separation based on colors or photometric redshift selection preclude further detailed weak lensing analyses. 

Combining the weak lensing analyses with background redshift distributions promises multiple useful applications. For example, \citet{Hu02} explains the idea of tomographic analyses based on a well-determined redshift distribution for the background sources. \citet{Medezinski11} and \citet{Kelly14} also suggest a cosmological test based on the reduced shear as a function of redshift for a modest well-chosen set of clusters (see also \citealp{Taylor07}). In particular, \citet{Kelly14} derive the weak lensing tomographic signal as a function of photometric redshift bins, and discuss the potential test for cosmological models. \citet{dellAntonio20} use simulated observations combining weak lensing measurements and extensive redshift surveys with an instrument like Subaru/Prime Focus Spectrograph (PFS) to demonstrate that tomographic analysis based on spectroscopic redshifts holds promise for the application of this independent geometric cosmological test.

Spectroscopic weak lensing tomography (hereafter spectrotomography) combines the weak lensing with dense spectroscopy to derive weak lensing signal for spectroscopically identified background galaxies, and tomographic lensing signal as a function of redshift of background galaxies. \citet{dellAntonio20} pioneered the power of this approach by applying it to the rich cluster, A2029. In their study of A2029, \citet{dellAntonio20} combine DECam imaging with a redshift survey of 2256 galaxies within $23'$ of the center of A2029 ($z = 0.078$). Based on the 1519 background galaxies, they detect a shear signal at $> 3.9\sigma$ that is consistent with the X-ray and dynamical mass of the cluster ($9 \times 10^{14}\rm{M}_\odot$, \citealp{Sohn19}). They also demonstrate that there is no contaminating signal from the 597 cluster members with spectroscopic redshifts.

Here we extend this spectrophotometric approach to two additional systems, A1767 and A2065. This study is part of an extended project to build the platform for spectrotomographic analyses, extending the work of \citet{dellAntonio20} and followed by Wright et al. (2026, in preparation). These two clusters also have extensive spectroscopy from the MAssive Cluster Survey with MMT/Hectospec \citep{Sohn20, Park26}. Additionally, they were observed with the wide-field imager Subaru/Hyper Suprime-Cam (HSC). In principle, Subaru/HSC is a more sensitive instrument than DECam, despite having a slightly smaller field of view. Furthermore, Subaru/HSC is mounted on a larger telescope at a site with better image quality. Comparison between the DECam imaging results and the Subaru/HSC analysis enables investigation of subtle systematics in deriving the shear for relatively bright ($r \lesssim 20.5$) background galaxies. Understanding these systematics is important for the analysis of larger future datasets.

Combining the redshift samples and the weak lensing shear measurements, we describe the detection of the tomographic signal for A1767 and A2065. We discuss the Subaru archival imaging data in Section \ref{sec:photometry} and the spectroscopy in Section \ref{sec:spectroscopy}. We then examine the ellipticity distributions of the background sources and cluster members and derive the first weak lensing maps based on sources with background redshifts alone in Section \ref{sec:wlshear}. We describe the weak lensing tomographic signal and its significance and discuss the subtle limitations in Section \ref{sec:tomog}. We preview future extensions of this work in Section \ref{sec:future} and we conclude in Section \ref{sec:conc}. We use Planck cosmology \citep{PlanckCollaboration16} with $H_{0} = 67.74$ km s$^{-1}$ Mpc$^{-1}$ and $\Omega_{m} = 0.3089$, and $\Omega_{\Lambda}$ = 0.6911 throughout the paper.

\section{SUBARU/HSC PHOTOMETRY AND SHAPES} \label{sec:photometry}

We derive the weak lensing signals for two massive clusters A1767 ($z = 0.071$) and A2065 ($z = 0.073$) in the local universe. These two clusters are low redshift targets for studying spectroscopic weak lensing tomography that combines lensing based on the Subaru imaging and MMT/Hectospec spectroscopy (see Section \ref{sec:spectroscopy}). We summarize the physical properties of the target clusters, including redshift, $M_{200}$ from the caustic technique \citep{Sohn20} in Table \ref{tab:summary}. 

\begin{table*}[h!]
\caption{The Physical Properties of A1767 and A2065}
\label{tab:summary}
\centering         
\begin{tabular}{c c c c c c}
\hline\hline              
ID & R.A. (J2000) & Decl. (J2000) & z & R$_{200, caustic}$ (Mpc) & M$_{200, Caustic}$ ($10^{14}$ M$_{\odot}$) \\         
\hline                  
A1767 & 13:36:09.16 & 59:11:15.88 & 0.0712 & $1.70 \pm 0.07$ &  $5.97 \pm 0.45$ \\
A2065 & 15:22:24.16 & 27:41:50.68 & 0.0731 & $2.03 \pm 0.08$ & $10.19 \pm 1.05$ \\
\hline
\end{tabular}
\end{table*}

We use archival HSC imaging as the source of photometry and shapes of galaxies. HSC, the wide-field optical imaging camera installed on the Subaru 8.2m telescope, has a 1.5 deg diameter field of view \citep{Miyazaki18}. HSC imaged A1767 and A2065 in the HSC-$i2$ (hereafter, $i$) band with a small dither to fill gaps between the CCDs. There were five exposures for A1767 and ten for A2065. The first five exposures were short (30 sec) for both clusters. For A2065, there were five additional longer (200 sec) exposures. The HSC images were taken under good seeing conditions: $0.6'' - 0.8''$ for A1767 and $0.45'' - 0.62''$ for A2065. 

We retrieved the processed catalogs from the HSC Legacy Archive (HSCLA; \citealp{hscla}), which were generated from coadded images processed with the standard \texttt{hscPipe} v.8 pipeline \citep{Bosch18,Aihara22}. We briefly describe the data reduction process here, and we refer to \citet{Bosch18} and \citet{Aihara22} for the full descriptions. Individual images were reduced according to the standard instrumental signature removal processes, including overscan, bias, dark, flat-fielding, bad-pixel masking, and correction for linearity, cross-talk, and brighter-fatter effects \citep{Antilogus14, Coulton18}. A global and local sky background subtraction was also performed.  

The \texttt{hscPipe} uses the PSF (Point Spread Function) models constructed based on \texttt{PSFEx} \citep{Bertin13}. \texttt{PSFEx} first selects candidate stars based on sizes of objects in each CCD, and models PSF focusing on the central $20 \times 20$ pixels of the PSF candidate stars. \texttt{PSFEx} constructs independent PSF models for each CCDs and fits a second-order polynomial to interpolate between stars distributed across the CCD. 

Finally, \texttt{hscPipe} derives photometry, including source position, flux, size, and shape measurements. We first obtain the cModel magnitude and its uncertainty, and extendedness through HSCLA. We also obtain the PSF-corrected shapes of sources derived based on the HSM re-Gaussianization method \citep{Hirata03, Hirata04, Mandelbaum05}. In this technique, both the observed source image and the local PSF are first approximated by elliptical Gaussian models using adaptive moments. The non-Gaussian components of the PSF are then accounted for by subtracting a correction term (i.e., ``re-Gaussianization") from the observed image, effectively reconstructing a Gaussianized galaxy profile prior to PSF convolution. We thus collect the PSF-corrected shear measurements (i.e., \textit{hsmshaperegauss$\_$e1} and \textit{hsmshaperegauss$\_$e2}). Here, the component $e_{1}$ represents elongation aligned with the $x-$axis (or perpendicular to the $y-$axes), while $e_{2}$ captures elongation at $45^{\circ}$ to these axes.

\begin{figure}
\centering
\includegraphics[scale=0.38]{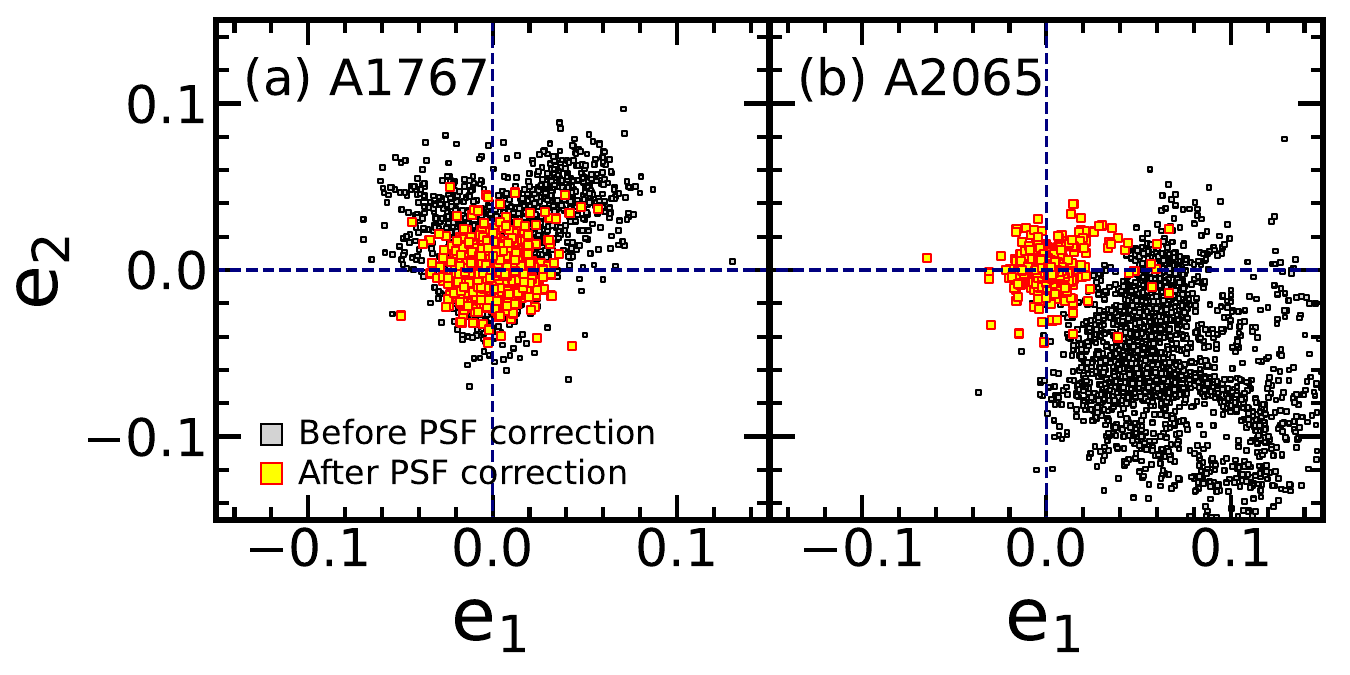}
\caption{Ellipticity distributions for point sources in (a) A1767 and (b) A2065 fields. Black and yellow squares show raw ellipticities and ellipticities after the PSF correction, respectively. }
\label{fig:ps}
\end{figure}

We test the PSF-corrected ellipticity measurements by checking the ellipticities of point-like sources. We select point-like sources with `extendedness'$ = 0$ and within $19 < i_{cModel} < 22$. Figure \ref{fig:ps} shows the ellipticity distributions of point-like sources. We obtain HSM adaptive secondary moments (i.e., $I_{xx}, I_{xy}$, and $I_{yy}$) and compute the ellipticity components (black points in Figure \ref{fig:ps}): 
\begin{equation}
e_{1} = \frac{I_{xx} - I_{yy}}{I_{xx} + I_{yy}}, 
\end{equation}
and
\begin{equation}
e_{2} = \frac{2 I_{xy}}{I_{xx} + I_{yy}}. 
\end{equation}
We then compute the PSF model ellipticities for each object, as determined by \texttt{hscpipe}. The yellow squares display the residuals for point-like sources. After the PSF correction, the mean ellipticities for point-like sources in the A1767 and A2065 fields are $0.001 \pm 0.01$. The concentration around zero ellipticity indicates that the PSF model is constructed, reasonably.

\section{SPECTROSCOPY}\label{sec:spectroscopy}

\begin{table*}[h!]
\caption{Cluster Redshift Samples}    
\label{tab:zsum}    
\centering                        
\begin{tabular}{l c c c c c c}      
\hline\hline               
Subsamples & \multicolumn{2}{c}{A2029$^{*}$} & \multicolumn{2}{c}{A1767}  & \multicolumn{2}{c}{A2065} \\
 & $ < R_{200}$ & $ < 2R_{200}$
 & $ < R_{200}$ & $ < 2R_{200}$
 & $ < R_{200}$ & $ < 2R_{200}$ \\         
\hline                      
Total redshifts                              & 1645 & 4462 & 1392 & 3979 & 2336 & 6258 \\
SDSS redshifts                               &  201 &  561 &  156 &  433 &  226 &  635 \\
MMT/Hectospec redshifts                      & 1379 & 3702 & 1235 & 3547 & 2103 & 5275 \\
Redshifts and shear measurements             & 1486 & 4017 & 1390 & 3965 & 2317 & 6172 \\
Spectroscopic members                        &  523 & 1062 &  323 &  589 &  617 &  841 \\
Spectroscopic background objects ($z > 0.1$) &  938 & 2948 & 1014 & 3264 & 1621 & 5038 \\
\hline                                  
\end{tabular}
\tabnote{$^{*}$ \citet{dellAntonio20}}
\end{table*}

\begin{table*}[h!]
\caption{Spectroscopic Redshifts in the A1767 field}    
\label{tab:z1767}    
\centering                        
\begin{tabular}{l c c c c}      
\hline\hline               
SDSS Object ID & R.A. & Decl. & z & z Source \\ 
\hline                      
1237655106770567239 & 198.86182 & 58.07962 & $0.11887 \pm 0.00004$ & SDSS \\
1237655106770567253 & 198.93708 & 58.16119 & $0.18958 \pm 0.00006$ & SDSS \\
1237655106770633026 & 199.34187 & 58.06649 & $0.19575 \pm 0.00004$ & SDSS \\
1237655106770568076 & 199.06066 & 58.15136 & $0.50106 \pm 0.00018$ & SDSS \\
1237655106770633432 & 199.39641 & 58.11600 & $0.54974 \pm 0.00011$ & SDSS \\
\hline
\end{tabular}
\tabnote{The entire table is available in machine-readable form in the online journal. Here, a portion is shown for guidance regarding its format.}
\end{table*}

\begin{table*}[h!]
\caption{Spectroscopic Redshifts in the A2065 field}    
\label{tab:z2065}    
\centering                        
\begin{tabular}{l c c c c}      
\hline\hline               
SDSS Object ID & R.A. & Decl. & z & z Source \\ 
\hline                      
1237662223546712109 & 228.58886 & 29.98727 & $0.54292 \pm 0.00016$ & SDSS \\
1237662223546778191 & 228.77192 & 29.96899 & $0.73603 \pm 0.00017$ & SDSS \\
1237662223547236476 & 229.92786 & 29.51339 & $0.10043 \pm 0.00001$ & SDSS \\
1237662223546974312 & 229.20159 & 29.76472 & $0.09830 \pm 0.00002$ & SDSS \\
1237662223547105492 & 229.57447 & 29.66332 & $0.24195 \pm 0.00005$ & SDSS \\
\hline                                  
\end{tabular}
\tabnote{The entire table is available in machine-readable form in the online journal. Here, a portion is shown for guidance regarding its format.}
\end{table*}

A1767 and A2065 are among nine clusters in the MACH (MAssive Clusters with Hectospec) survey \citep{Park26}. MACH is a complete sample of clusters in the northern sky (Decl. $> 5^{\circ}$) with dynamical masses, computed using the caustic technique \citep{Diaferio97, Diaferio99}, greater than $5 \times 10^{14}~\Msun$, in the redshift range $0.07 < z < 0.09$.  

We began by collecting redshifts for galaxies from SDSS spectroscopy, which is complete for galaxies brighter than $r=17.77$ \citep{Strauss02}. There are 148 and 206 objects with SDSS spectra within $R_{200}$ radius for A1767 and A2065, respectively. The SDSS spectra are acquired through a $3''$ diameter fiber. The typical redshift uncertainty ($c\Delta z$) for these spectra is $\sim 13~\kms$. We additionally collect two and fourteen unique redshifts for A1767 and A2065, respectively, from the Dark Energy Spectroscopic Instrument (DESI, \citealp{DESI}), with redshift uncertainties comparable to those of the SDSS redshifts.

Between 2018 and 2023 we conducted denser, more complete redshift surveys of A1767 and A2065 using MMT/Hectospec \citep{Fabricant05} following the approach of \citet{Sohn17a} and \citet{Sohn19}. The survey targets are galaxies with $r_{cModel,0} \leq 21.3$ and $r_{fiber} \leq 22$ in the SDSS photometric galaxy catalog. The fiber magnitude selection excludes low surface brightness galaxies. There is no color selection in these surveys. 

The 270 line mm$^{-1}$ grating of Hectospec yields spectra with a $6.2~\mathrm{\mathring{A}}$ covering the wavelength range $3700 < \lambda~(\mathrm{\mathring{A}}) < 9150$. For each Hectospec field, we take three 1200-second exposures to enable cosmic ray removal.

We reduced the Hectospec spectra using the standard pipeline, HSRED v2.0. We used the cross-correlation tool, RVSAO \citep{Kurtz98} to derive redshifts. RVSAO yields the cross-correlation score (R$_{XC}$). Following our previous Hectospec surveys and analyses, we selected reliable redshifts with $R_{XC} > 3$ (see Figure 5 of \citealp{Sohn21a}). The typical redshift uncertainty in a Hectospec redshift is $\sim 39~\kms$. For a small number of objects with both Hectospec and SDSS/DESI redshifts (one in A1767 and 60 in A2065), we chose the Hectospec redshifts. There are now a total of 1392 (3979) redshifts in the A1767 field and 2336 (6258) in the A2065 field within $R_{200}$ (and 2$R_{200}$).

There is a small offset between Hectospec and SDSS/BOSS redshifts. The Hectospec redshifts are slightly lower: $c\Delta(z_{SDSS/BOSS} - z_{Hecto}) / (1 + z_{Hecto}) = 42~\kms$. \citet{Kim25} explore this issue in detail and find that the offset results from a zero-point offset in the MMT templates. However, we do not correct for this offset because it does not affect any of the analysis here.

\begin{figure}
\centering
\includegraphics[scale=0.25]{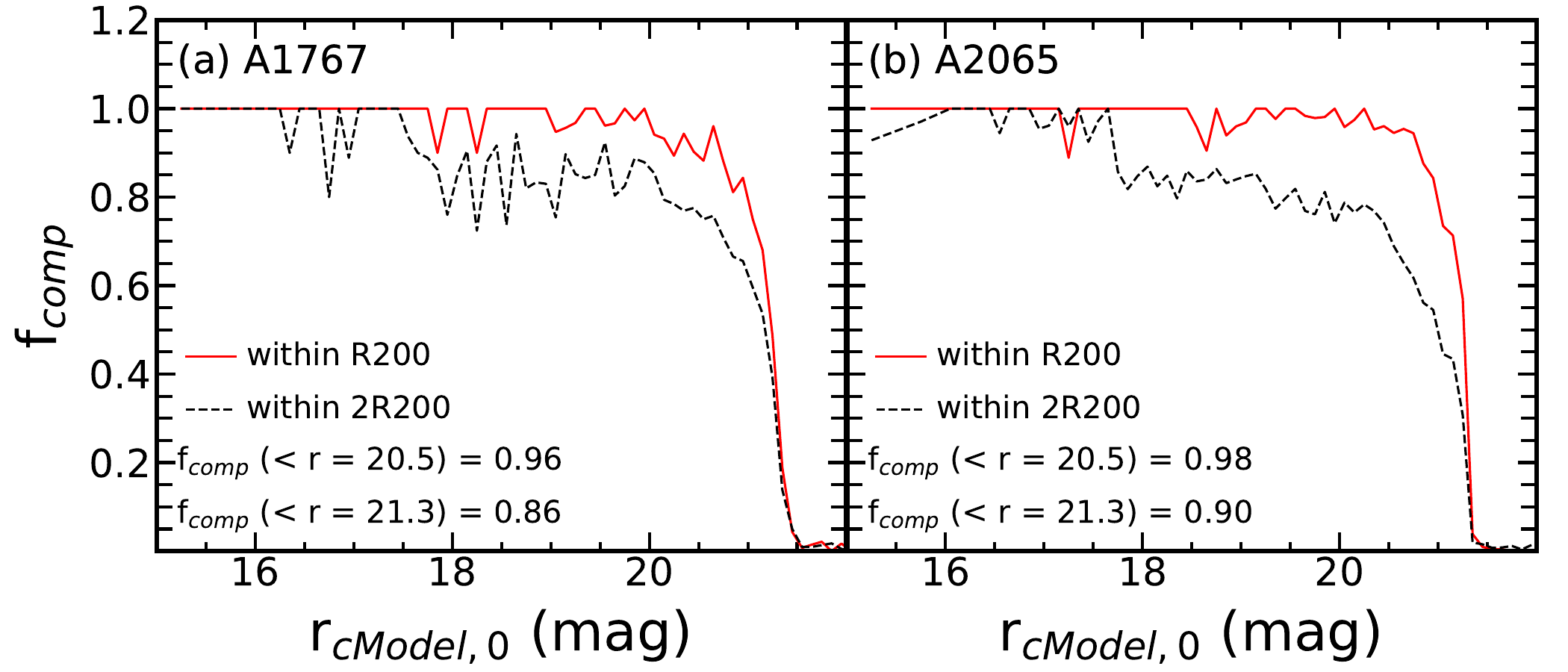}
\caption{Spectroscopic survey completeness for galaxies within the $R_{200}$ from the center of (a) A1767 and (b) A2065 as a function of $r-$band magnitude, respectively. } 
\label{fig:complete}
\end{figure}

Figure \ref{fig:complete} shows the completeness of each cluster survey within $30'$. To a limiting magnitude $r = 20.5$, the completeness of both surveys is $> 97\%$. A1767 and A2065 are among the best-sampled clusters in their redshift range. Among the galaxies with redshifts, more than 98.5\% of them have shapes derived from HSC imaging. Table \ref{tab:zsum} lists the number of spectroscopic redshifts in various subsamples within the A1767 and A2065 fields. Tables \ref{tab:z1767} and \ref{tab:z2065} list the redshifts for individual galaxies in the A1767 and A2065 fields, respectively. 

We use the caustic technique \citep{Diaferio97, Diaferio99, Serra13} to identify cluster members and to separate them from foreground and background galaxies. The caustic technique derives the escape velocity profile as a function of the projected clustercentric distance. Galaxies that fall within the caustics in phase space are cluster members; those outside are foreground or background galaxies. Table \ref{tab:zsum} summarizes the number of spectroscopically confirmed cluster members, foreground galaxies, and background galaxies for both systems, alongside corresponding values for A2029 from \citet{dellAntonio20} for comparison.

\begin{figure*}
\centering
\includegraphics[scale=0.35]{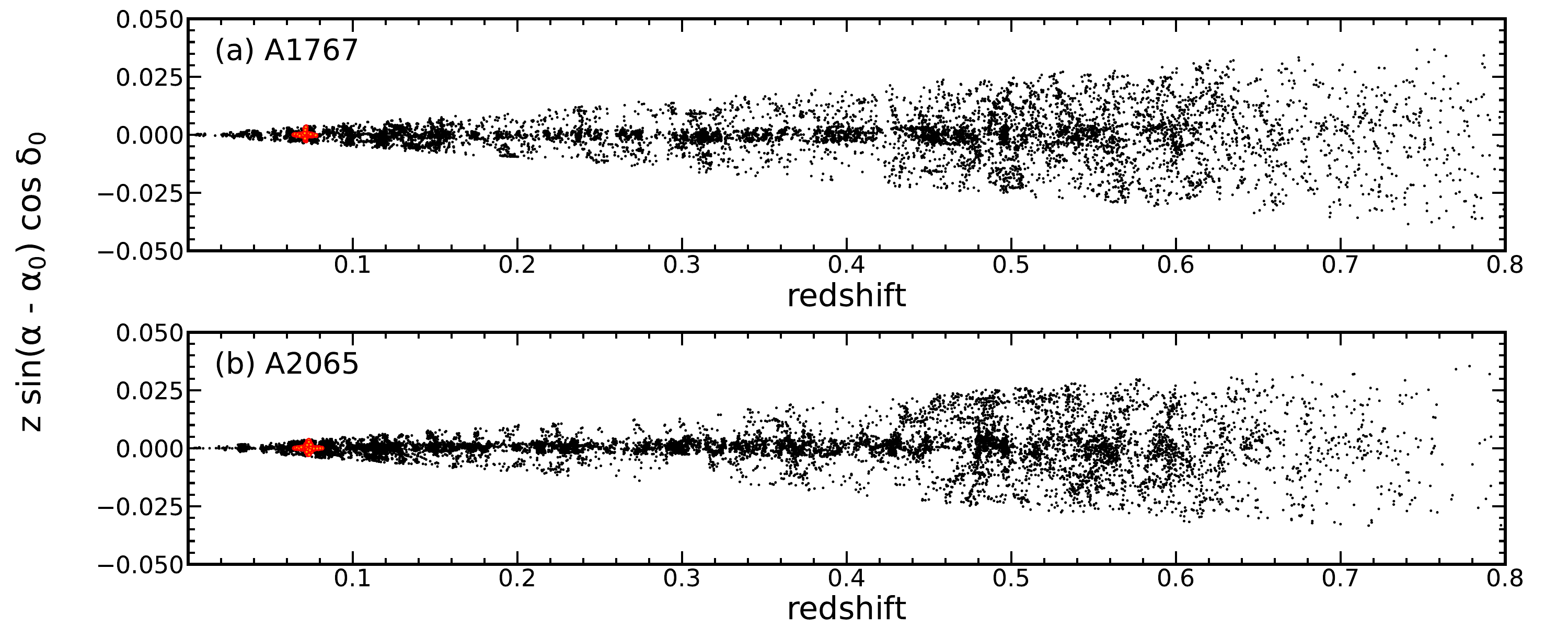}
\caption{Cone diagrams for (upper) the A1767 field and (lower) the A2065 field. Black dots are galaxies with spectroscopic redshifts, and red dots mark the spectroscopic members of the two clusters. }
\label{fig:cone}
\end{figure*}

Figure \ref{fig:cone} shows cone diagrams for A1767 and A2065, respectively. Black points represent galaxies with spectroscopic redshifts and red points indicate cluster members. In both fields, cluster members exhibit the expected patterns: line-of-sight elongation due to the finger-of-god effect and transverse elongation due to the infall region. More importantly, the cone diagrams reveal voids and large-scale structures along the line of sight. At $z \gtrsim 0.5$, these structures become poorly defined as the redshift survey becomes sparse.

\begin{figure}[h!]
\centering
\includegraphics[scale=0.25]{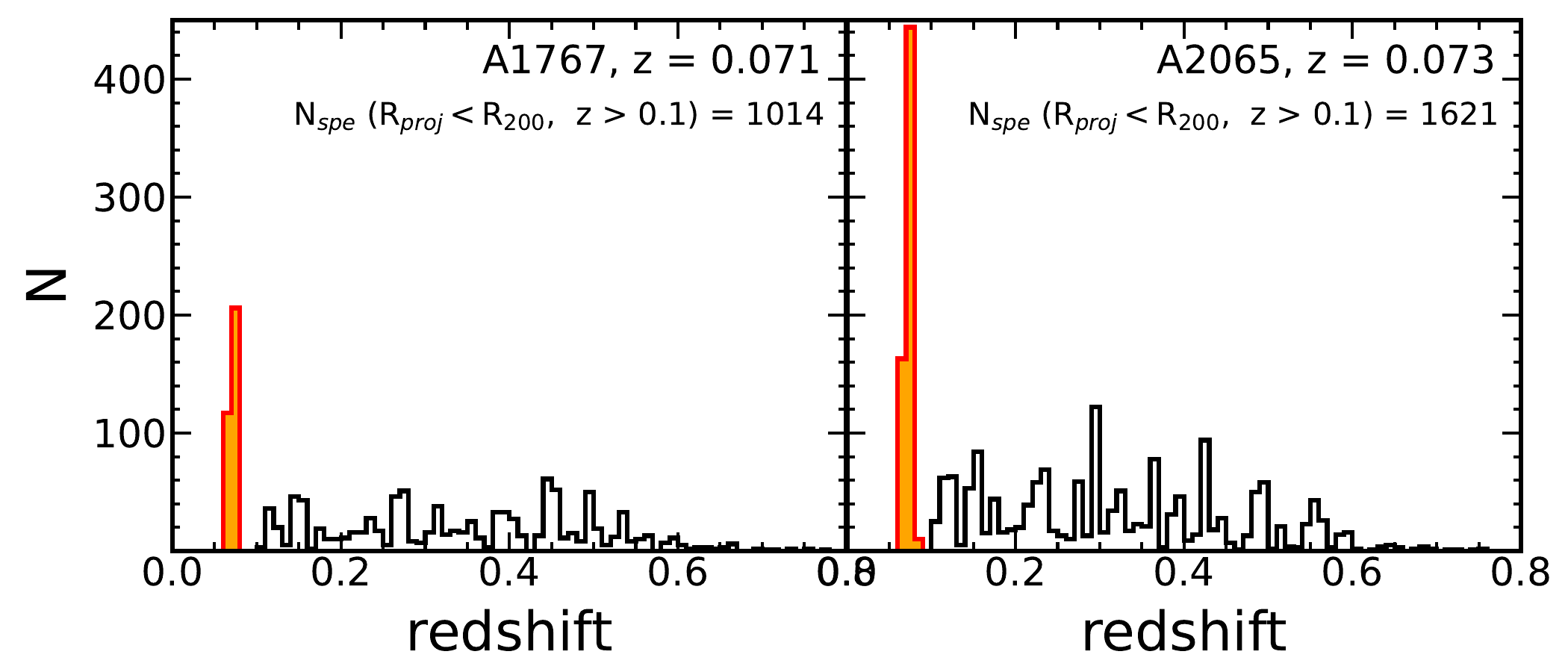}
\caption{Redshift distribution for (a) the A1767 field and (b) the A2065 field. The red histogram highlights cluster members. }
\label{fig:zhist}
\end{figure}

Figure \ref{fig:zhist} summarizes the redshift distributions for both clusters. The red outline highlights the cluster peak in each histogram. The histograms show the sequence of background peaks that correspond to the dense structures in the cone diagrams. We use background objects with $z > 0.1$ to explore the spectrotomographic signal.

\section{THE WEAK LENSING SHEAR: ELLIPTICITY DISTRIBUTIONS AND MASS MAPS} \label{sec:wlshear}

We use Subaru HSC shapes as the basis for weak lensing maps and tomographic lensing measurements. We particularly use HSM ellipticity rather than shear because our goal is signal detection rather than precise mass estimation. The galaxies in the spectroscopic tomographic analyses are generally bright and well-resolved in HSC imaging. Thus, using ellipticities directly allows for more straightforward modeling and signal-to-noise estimation, at the cost of transforming reference shear estimates back to ellipticities given the cluster masses. While multiplicative and additive biases in this transformation could affect comparisons with theoretical predictions, these biases based on HSC imaging are at the sub-percent level ($\sim 0.005$, \citealp{Mandelbaum18}), well below the uncertainties in the mean tomographic signal ($\sim0.01$). The agreement between the data and the predictions is therefore insensitive to the choice between shear and ellipticity. We present all of the results in terms of ellipticities rather than converting the PSF-corrected ellipticities into shear measurements.

The ellipticity distributions for spectroscopically identified background galaxies and for cluster members provide tests of systematics and evidence for the shear signal detection. First, the distributions of $e_{1}$ and $e_{2}$ of all galaxies should be identical in the absence of systematics that may occur due to the PSF subtraction, shape measurements, and intrinsic alignments. Second, the distributions of tangential (i.e., $+$, E-mode) and cross (i.e., $\times$, B-mode) components for the cluster members should be identical because their shapes are not affected by the gravitational field of the cluster. Third, in the presence of a lensing signal, the tangential ellipticity distribution of the background galaxies should be significantly offset from the distribution of their cross ellipticity relative to the cluster center. Furthermore, the cross ellipticity signals of the background galaxies should be consistent with zero. 

We first examine the systematics based on the ellipticity distributions of cluster members and the background galaxies (Section \ref{sec:systematics}). We then examine the detection of shear signal based on the spectroscopically identified background galaxies (Section \ref{sec:detection}). In Section \ref{sec:maps}, we construct the weak lensing signal maps based on the photometric sample of background galaxies and on the much smaller set of spectroscopically identified background galaxies to demonstrate the significance of the detection.

\subsection{Examining Systematics in Ellipticity Measurements} \label{sec:systematics}

We first examine the ellipticity measurements to test the absence of systematic effects based on two galaxy samples (Table \ref{tab:zsum}): cluster members identified via the caustic technique, and spectroscopically confirmed background galaxies with $z > 0.1$ in each cluster field. 

\begin{figure}[h!]
\centering
\includegraphics[scale=0.35]{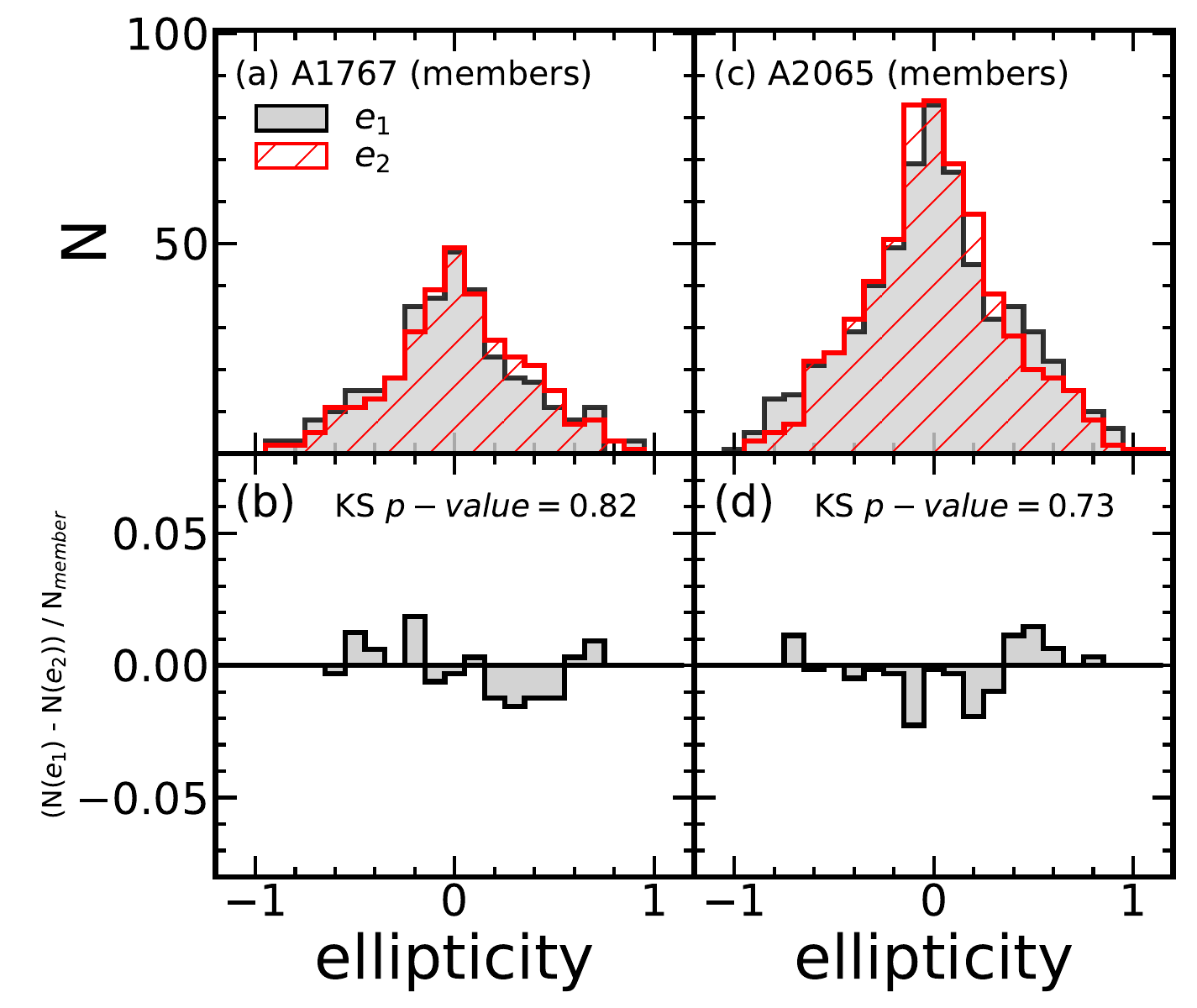}
\caption{(a) The distributions of $e_{1}$ (the black filled) and $e_{2}$ (the red hatched) of spectroscopically identified members within $R_{200}$ from the A1767 center. (b) The normalized difference between $e_{1}$ and $e_{2}$ distributions. (c-d) Same as (a-b), but for A2065 member galaxies. }
\label{fig:e1e2_members}
\end{figure}

Azimuthal symmetry around the center of each cluster should produce indistinguishable distributions of the ellipticities $e_{1}$ and $e_{2}$ for both cluster members and background galaxies in the absence of systematics. The upper panels of Figure \ref{fig:e1e2_members} show histograms of $e_{1}$ (the black filled) and $e_{2}$ (the red hatched) distributions for cluster members within $R_{200}$ of the center of A1767 (left) and A2065 (right). The lower panels display the normalized difference $(N(e_{1})- N(e_{2})) / N_{member}$ as a function of ellipticity. Throughout the ellipticity range, the differences are within $1\sigma$. We also applied the Kolmogorov-Smirnov (KS) test to these distributions, and the resulting $p-$values for the two clusters are $> 0.73$, indicating that the null hypothesis that $e_{1}$ and $e_{2}$ distributions are driven from the same parental distribution is not rejected. The test with the $e_{1}$ and $e_{2}$ distributions suggests that the cluster members do not show preferential alignment. 

\begin{figure}[h!]
\centering
\includegraphics[scale=0.32]{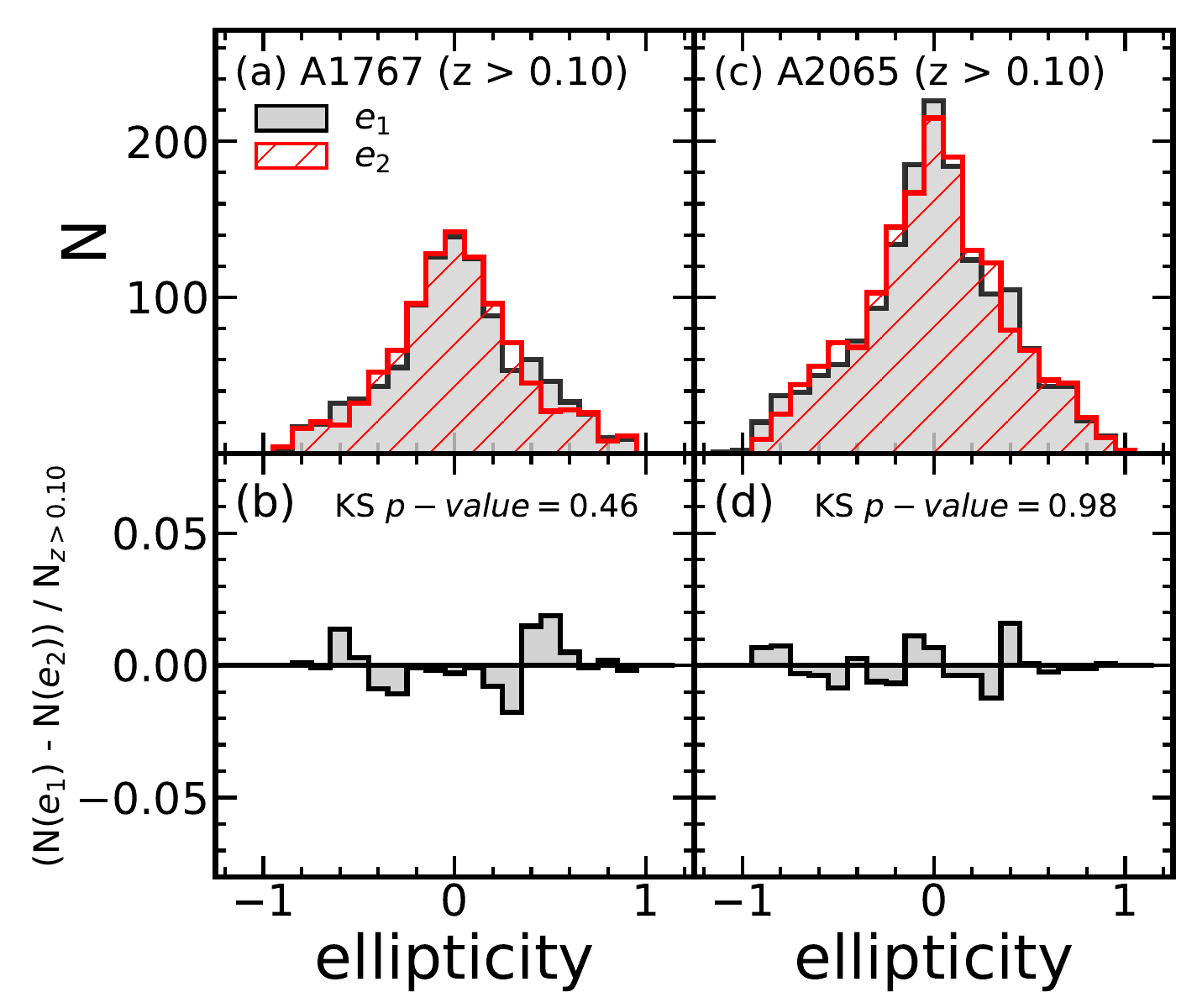}
\caption{(a) The distributions of $e_{1}$ (the black filled) and $e_{2}$ (the red hatched) of spectroscopically identified background galaxies (i.e., $z > 0.1$) within $R_{200}$ from the A1767 center. (b) The normalized difference between $e_{1}$ and $e_{2}$ distributions. (c-d) Same as (a), but for A2065 member galaxies. }
\label{fig:e1e2_background}
\end{figure}

Similarly, Figure \ref{fig:e1e2_background} shows distributions of $e_{1}$ and $e_{2}$ (the upper panels) and the normalized difference ($(N(e_1)- N(e_2))/N_{\rm background}$, the lower panels) for the background galaxies in the two cluster fields. The differences are insignificant throughout the ellipticity range. The high $p-$values from the KS tests (0.46 for A1767 and 0.98 for A2065) also suggest that the $e_{1}$ and $e_{2}$ distributions are not significantly different. Therefore, there is no significant systematic orientation of the background sources that might contaminate the lensing signal.

We also examine the tangential ($e_{tan}$) and cross ($e_{cross}$) ellipticity distributions of cluster members. Because cluster members do not experience image distortions due to the gravitational fields of the target clusters, their $e_{tan}$ and $e_{cross}$ distributions should not show any systematic differences. 

\begin{figure}
\centering
\includegraphics[scale=0.35]{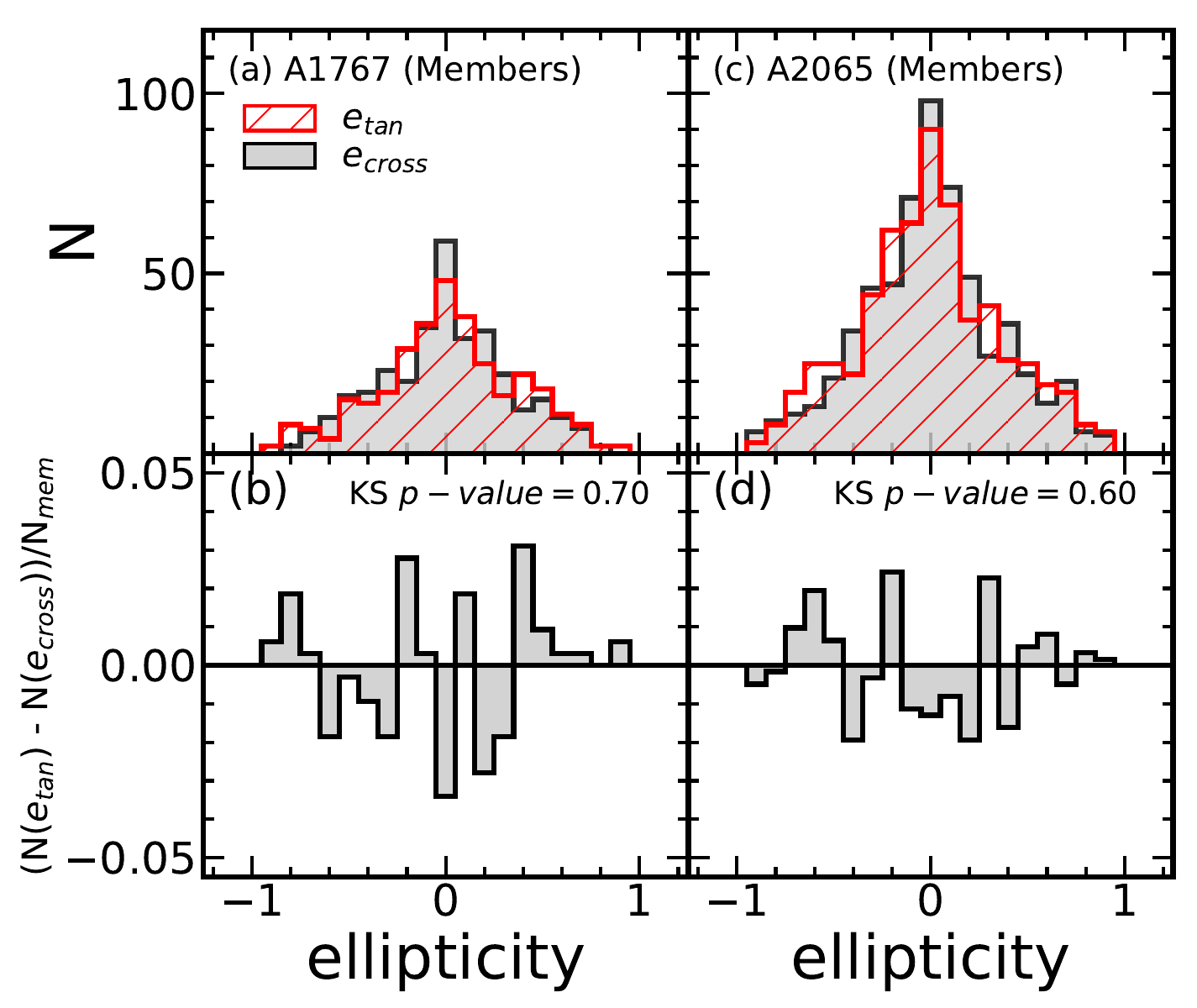}
\caption{(a) Distributions of $e_{tan}$ (the red hatched) and $e_{cross}$ (the black filled) of spectroscopically identified member galaxies within $R_{200}$ for A1767. (b) Difference between $e_{tan}$ and $e_{cross}$ distributions for A1767. (c-d) Same as (a-b), but for A2065 spectroscopically identified members.}
\label{fig:etec_member} 
\end{figure}

Figure \ref{fig:etec_member} compares the raw and normalized distributions of $e_{tan}$ (the red hatched) and $e_{cross}$ (the black filled) for members of A1767 and A2065 within $R_{200}$ of the cluster center. There is no significant difference between the $e_{tan}$ and $e_{cross}$ distributions for members in either cluster, consistent with the expectation that cluster members do not experience lensing distortion from the cluster potential. The KS test $p-$values for A1767 and A2065 ellipticity distributions are 0.70 and 0.60; the $e_{tan}$ and $e_{cross}$ distributions for the cluster members are not significantly distinctive.

\subsection{Weak Lensing Signal Detection based on Spectroscopic Background Galaxies}\label{sec:detection}

We next demonstrate the weak lensing signal detection based only on spectroscopically identified background galaxies. Unlike the cluster members (as shown in Figure \ref{fig:etec_member}), the distribution of $e_{tan}$ should be offset toward larger (positive) compared to the distribution of $e_{cross}$, which is consistent with zero, in the presence of lensing. 

\begin{figure}[h!]
\centering
\includegraphics[scale=0.34]{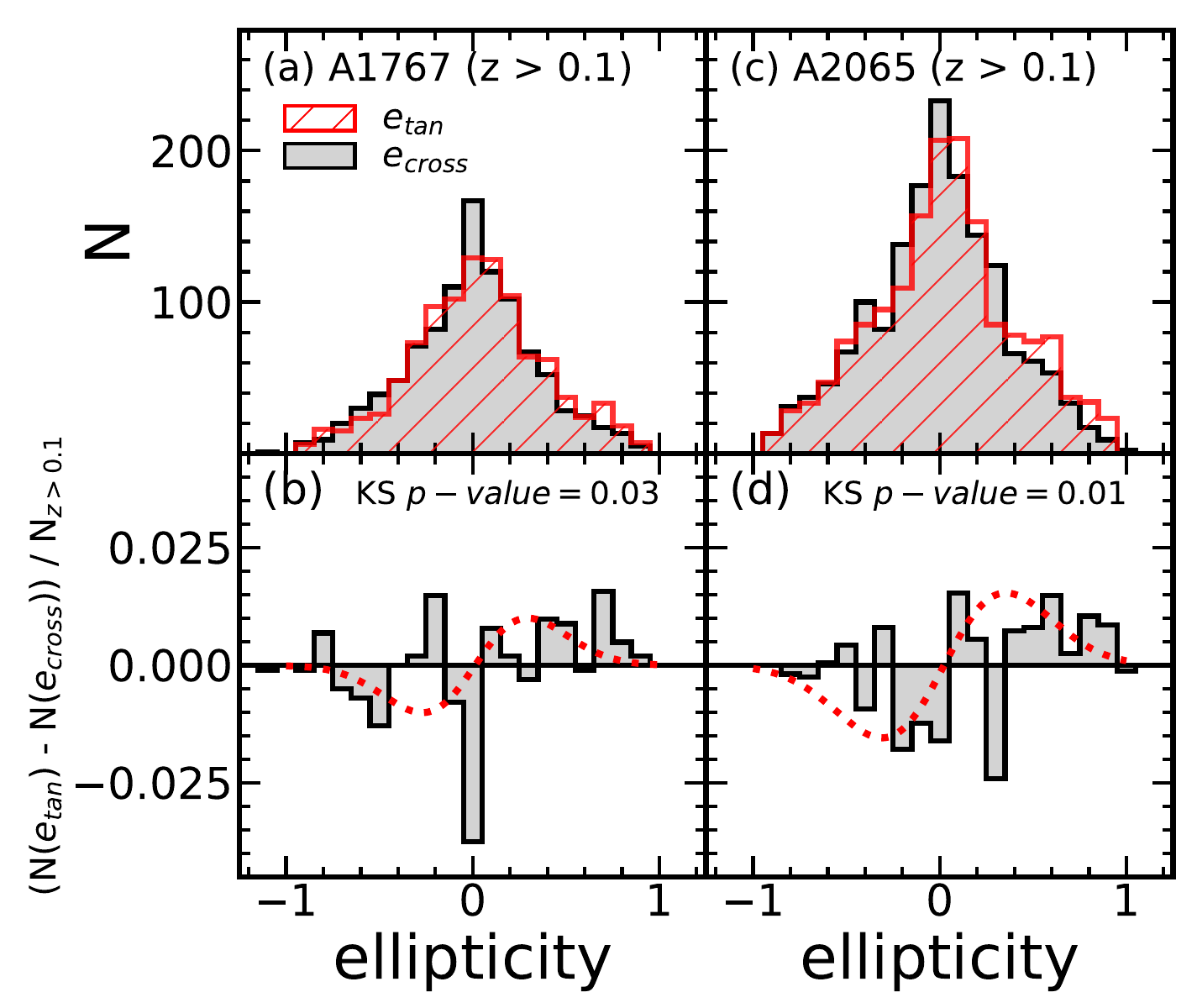}
\caption{(a) Distributions of $e_{tan}$ (the red hatched) and $e_{cross}$ (the black filled) of spectroscopically identified background ($z > 0.1$) galaxies within $R_{200}$ from the A1767 center. (b) Difference between $e_{tan}$ and $e_{cross}$ distributions for A1767. (c-d) Same as (a-b), but for spectroscopically identified background galaxies in the A2065 field. }
\label{fig:etec_background} 
\end{figure}

Figure \ref{fig:etec_background} shows the distributions of $e_{tan}$ (red hatched histograms) and $e_{cross}$ (black filled histograms) for background galaxies within $R_{200}$ from the centers of A1767 and A2065, respectively. The distributions of $e_{tan}$ for background galaxies are skewed distribution toward positive ellipticities. The low $p-$values (i.e., 0.03 for A1767 and 0.01 for A2065) from the K-S tests also indicate that the null hypothesis that $e_{tan}$ and $e_{cross}$ distributions for the background galaxies are driven from the same parental distribution can be rejected. 

For the more massive system, A2065, the expected shift of the $e_{tan}$ histogram toward larger ellipticity is unambiguous. A1767 also has a significant offset, but the distribution is noisier than for A2065. Because the number of background galaxies is similar for the two clusters, the noise in the shift in the field of A1767 is probably driven by its lower mass. 

The lower panels of Figure \ref{fig:etec_background} show the normalized difference between $e_{tan}$ and $e_{cross}$ distributions (i.e., $(N(e_{tan})- N(e_{cross}))/N_{\rm background}$) that highlight the shift of the $e_{tan}$ histogram toward larger ellipticity. The dashed curves show the expected signal for the cluster mass based on the caustic technique \citep{Sohn20}. Because the caustic (dynamical) masses and lensing signals are completely independent, the match between the data and the predicted amplitude of the signals is impressive.

\subsection{Weak Lensing Maps} \label{sec:maps}

We next construct weak lensing mass maps ($\kappa$) following \citet{KS93}:
\begin{eqnarray}
\kappa(\vec{\theta}) = \int d^2 \phi \gamma_{t} (\vec{\phi}; \vec{ \theta}) Q(|\vec{ \phi|}),
\end{eqnarray}
where $\vec{\theta}$ and $\vec{\phi}$ are celestial coordinates and $\gamma_{t}$ is the tangential component of the shear $\vec{\gamma}$ with respect to $\vec{\theta}$. The weight function $Q(\theta)$ is defined as:
{\small
\begin{eqnarray}
Q(\theta) = \begin{cases} 
\dfrac{1}{\pi \theta^2}\left[ 1-\left( 1+\dfrac{\theta^2}{\theta_{G}^2} \right)\exp\left(-\dfrac{\theta^2}{\theta_{G}^2}\right) \right] & \text{for~} \theta < \theta_{o} \\
0 & \text{elsewhere.} 
\end{cases}
\end{eqnarray}
}
This weight function operates as a Gaussian smoothing kernel applied to the $\kappa$ map, where $\theta_{G}$ corresponds to the Gaussian smoothing scale and $\theta_{o}$ corresponds to the truncation scale. We evaluate the $\kappa$ field on a grid of $0.15'$ pixels in the flat-sky approximation and apply a Gaussian smoothing with $\theta_{G} = 3.0'$ and $\theta_{o}=15'$. The convergence map is computed using the \texttt{gamsky2kap$\_$gauss$\_$v1.3.f} package\footnote{http://sci.nao.ac.jp/MEMBER/hamana/OPENPRO/gamsky2kap$\_$gauss$\_$v1.3.f} \citep{Hamana15, Hamana20}. The noise map is estimated as the $1\sigma$ standard deviation derived from 100 realizations generated by randomly resampling the orientations of background galaxies. In practice, galaxy ellipticities are used in place of estimated shear for $\gamma_{t}$; however, this choice does not affect the results, because our analysis relies on the signal-to-noise (S/N) maps rather than the absolute values of $\kappa$.

We construct weak lensing maps using two catalogs: (1) a ``spec sample'' containing only galaxies with spectroscopic redshifts, and (2) a ``full sample'' containing all photometric galaxies. For the full sample, we apply only weak quality cuts to the total photometric catalog: $|e| < 2$ and $i-$band cModel magnitude brighter than 25, and extendedness greater than 0.1. Here, $e$ denotes the HSM ellipticity measurements used to derive the lensing signal. We note that a small fraction of galaxies have ellipticities exceeding unity, the conventional maximum value, due to corrections applied by the HSM algorithm. However, the number of such objects is negligible; there are less than three objects in the spectroscopic sample and fewer than 5\% of objects in the full sample across both clusters. Their inclusion has little impact on the computed lensing maps.

\begin{figure*}[h!]
\centering
\includegraphics[scale=0.6]{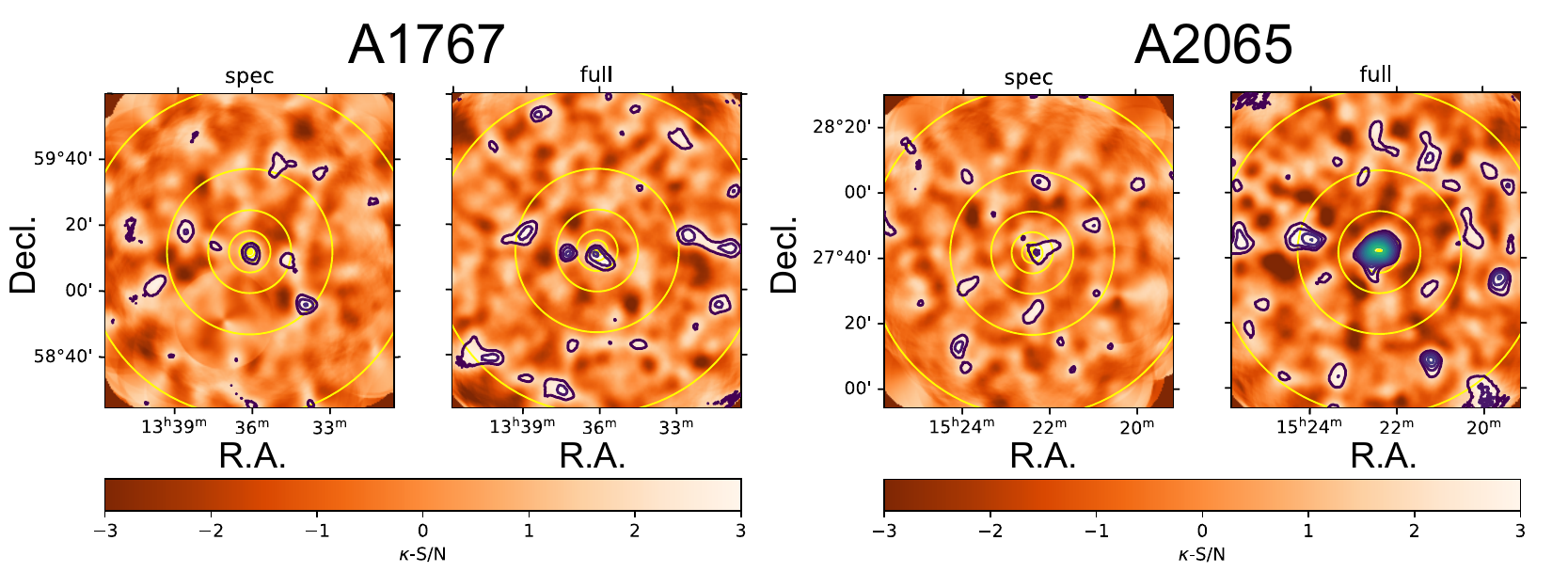}
\caption{Weak lensing significance ($\kappa$) maps for A1767 (left two panels) and A2065 (right two panels). For each cluster, the left panel shows the significance map constructed from spectroscopically identified background galaxies only, while the right panel shows the same map constructed from the full photometric background galaxy sample.} 
\label{fig:wlmaps}
\end{figure*}

Figure \ref{fig:wlmaps} shows the weak lensing significance maps for A1767 and A2065. For each cluster, the left panel displays the spectroscopic weak lensing map constructed from background sources with spectroscopic redshifts, and the right panel shows the map built from the full photometric background sample. A clear lensing signal is evident in both maps. As expected from the considerably smaller number of sources, the signal significance is lower in the spectroscopic maps, though its spatial extent remains unchanged.

For A1767, the signal of the spectroscopic weak lensing map has a peak significance of $4.0 \sigma$ compared with $4.3 \sigma$ for the photometric weak lensing map. The extent of the significant spectroscopic weak lensing signal is $1.5'$, which is comparable with the $1.5'$ extent of the full weak lensing map. The spectroscopic weak lensing map is based on a 0.3 galaxy per square arcmin background source density and demonstrates the promise of the technique when background redshift samples are more than an order of magnitude larger. The full sample is based on a 17.6 galaxy per square arcmin background source density.

We note that the full photometric sample for A1767 contains many poorly resolved galaxies due to the shallower depth and worse seeing of the images. Because the mass mapping algorithm does not account for the lower effective weight of these galaxies, the effective source density for A1767 is reduced significantly. This effect, combined with potential contamination from the wings of cluster galaxies affecting smaller background galaxies, may account for the relatively modest increase in detection significance of the photometric sample map compared to the spectroscopic sample map, where all galaxies are well-resolved and carry higher effective weight.

For A2065, the spectrotomographic signal has a peak significance of $3.1 \sigma$ compared with $14.0 \sigma$ for the full weak lensing map. The extent of the significant spectrotomographic signal is $1.3'$ compared with the $1.7'$ extent of the full weak lensing map. For the spectrotomographic map, the background source density is 0.49 galaxies per square arcmin; for the full map, the background source density is 35 galaxies per square arcmin. The baseline photometric observations for A2065 are significantly deeper than for A1767.

In summary, we construct weak lensing maps based only on background sources with redshifts in the fields of A1767 and A2065. These maps are the first spectrotomographic weak lensing maps. Eventually, much larger samples of background galaxies with spectroscopic redshifts will significantly increase the significance of these detections.

\section{TOMOGRAPHY}\label{sec:tomog}

We use the mean tangential ellipticity ($\langle e_{tan} \rangle$) to measure the weak lensing signal (rather than the calculated shear, as explained above in Section \ref{sec:wlshear}). For this analyses, we used 1014 and 1621 background ($z > 0.1$) galaxies within a projected radius of $R_{200}$ at the A1767 and A2065 fields, respectively. 

Figure \ref{fig:spectrotomography} displays $\langle e_{tan} \rangle$ as a function of the source redshift for sources with spectroscopic redshifts (the E-mode, red circles). As a measure of the significance of the signal we also show $\langle e_{cross} \rangle$ (the B-mode, black crosses). For sources with $z > 0.1$, the B-mode signal is generally consistent with zero within the uncertainties. In A1767, the E-mode signals exceed the B-mode signals at $0.1 < z < 0.4$, but this excess becomes negligible at higher redshifts. We suspect that the weak signal at higher redshifts is due to the limited number of spectroscopic background galaxies. In A2065, the tomographic E-mode signal significantly exceeds the B-mode across the full redshift range explored.

\subsection{The Expected Tomographic Signal} \label{sec:tomo_signal}

We construct model spectrotomographic signals based on cluster masses determined from the caustic technique: $M_{200} = 5.97 \times 10^{14}~\Msun$ for A1767 and $M_{200} = 1.02 \times 10^{15}~\Msun$ for A2065. We then compute the expected signal from a spherically symmetric NFW cluster (with concentration parameter, $c_{\rm NFW} = 4$) appropriate for clusters in the caustic mass range. For each galaxy in the spectroscopic sample, we calculate the critical density: 
\begin{eqnarray}
\Sigma_{crit} = {c_{s}^2\over 4\pi G}{D_S\over D_{L} D_{LS}},
\end{eqnarray}
where $c_{s}$ is the speed of light, $D_{L}$, $D_{S}$, and $D_{LS}$ are the angular diameter distances. Given the the projected distance $R=D_{L}\theta$ and the scaled distance $x = R_{c} / R_{200}$, we use the shear calculated from the analytic expressions in Equations 14-16 from \citet{Wright00} (we use $q(x)$ for their function $g(x)$ to remove confusion with the reduced shear $g$):
\begin{eqnarray}
\gamma(x) = r_{s} \delta_{c} \rho_{c} q(x), 
\end{eqnarray}
where $r_{s} = R_{200}/c$, $\rho_{c}$ is the critical density and
\begin{equation}
\delta_{c}= \frac{200}{3} \frac{c^{3}}{\ln (1+c) - c/(1+c)},
\end{equation}
where $c$ indicates the concentration parameter. The function $q(x)$ depends on the value of $x$: for $x < 1$,
\begin{eqnarray}
q(x) &=& \frac{8~\mathrm{arctanh}\left(\sqrt{\frac{(1-x)}{(1+x)}}\right)}{x^2\sqrt{1-x^2}} + \frac{4}{x^2} \ln \left(\frac{x}{2}\right) \nonumber \\
     && - \frac{2}{x^2 - 1} + \frac{4~\mathrm{arctanh}\left(\sqrt{\frac{(1-x)}{(1+x)}}\right)}{(x^2 - 1)\sqrt{1 - x^2}},
\end{eqnarray}
for $x=1$, 
\begin{eqnarray}
q(x) = {10\over 3} + 4 \ln (x/2),
\end{eqnarray}
and for $x>1$,
\begin{eqnarray}
q(x) &=& \frac{8~\mathrm{arctan} \left( \sqrt{\frac{(x - 1)}{(1 + x)}} \right)}{x^2 \sqrt{x^2 - 1}} + \frac{4}{x^2} \ln \left( \frac{x}{2} \right) \nonumber \\
     && - \frac{2}{x^2 - 1} + \frac{4~\mathrm{arctan} \left( \sqrt{\frac{(x - 1)}{(1 + x)}} \right)}{(x^2 - 1)^{3/2}}.
\end{eqnarray}

We also compute the convergence $\kappa(x)$ following \citet{Wright00}. The convergence $\kappa(x)$ is defined as
\begin{equation}
\kappa(x) = \Sigma(x) / \Sigma_{crit}. 
\end{equation}
Here $\Sigma(x)$ is defined as: 
\begin{equation}
\Sigma(x) = 2 r_{s} \delta_{c} \rho_{c} S(x), 
\end{equation}
where
\begin{equation}
S(x) = \frac{1}{x^{2}-1} \Big(1 - \frac{2}{\sqrt{(1-x^{2})}}\mathrm{arctanh}\frac{\sqrt{(1-x)}}{\sqrt{(1+x)}} \Big), 
\end{equation}
for $x < 1$,
\begin{equation}
S(1)=\frac{1}{3}, 
\end{equation}
and
\begin{equation}
S(x) = \frac{1}{x^{2}-1} \Big(1 - \frac{2}{\sqrt{x^{2}-1}} \mathrm{arctan} \Big(\frac{x-1}{\sqrt{x+1}} \Big) \Big), 
\end{equation}
for $x>1$. The observed reduced shear is $g = {\gamma \over (1-\kappa)}$.

To convert the reduced shear to ellipticity, we use the inverse of the weight prescription in \citet{Bosch19}:
\begin{equation}
e_{tan} = 2g (1 - \langle e^{2} \rangle),
\end{equation}
where $\langle e \rangle$ is the mean ellipticity of the galaxies. In principle, there is an additional component due to the uncertainty in the measured ellipticity. However, the galaxies with spectroscopic redshifts are both bright and well-resolved. Thus, the measurement errors in the HSC Subaru Strategic Program (SSP) data are negligible compared to the shape noise. Thus, we use the same weight for each galaxy. These equal weights simplify the calculation of the correction by allowing a global rather than a per-object correction. We use the mean ellipticity for the full spectrotomographic sample for both clusters $\langle e \rangle = 0.42$. The differences between the two fields amount to less than 1\%, and small differences in the ellipticity have a negligible effect on the curves on the scale of our measurements. We average the predicted ellipticities of all the galaxies in each redshift bin to predict the signal for the sample of galaxies with spectroscopic redshifts. 

\begin{figure*}[h!]
\centering
\includegraphics[scale=0.45]{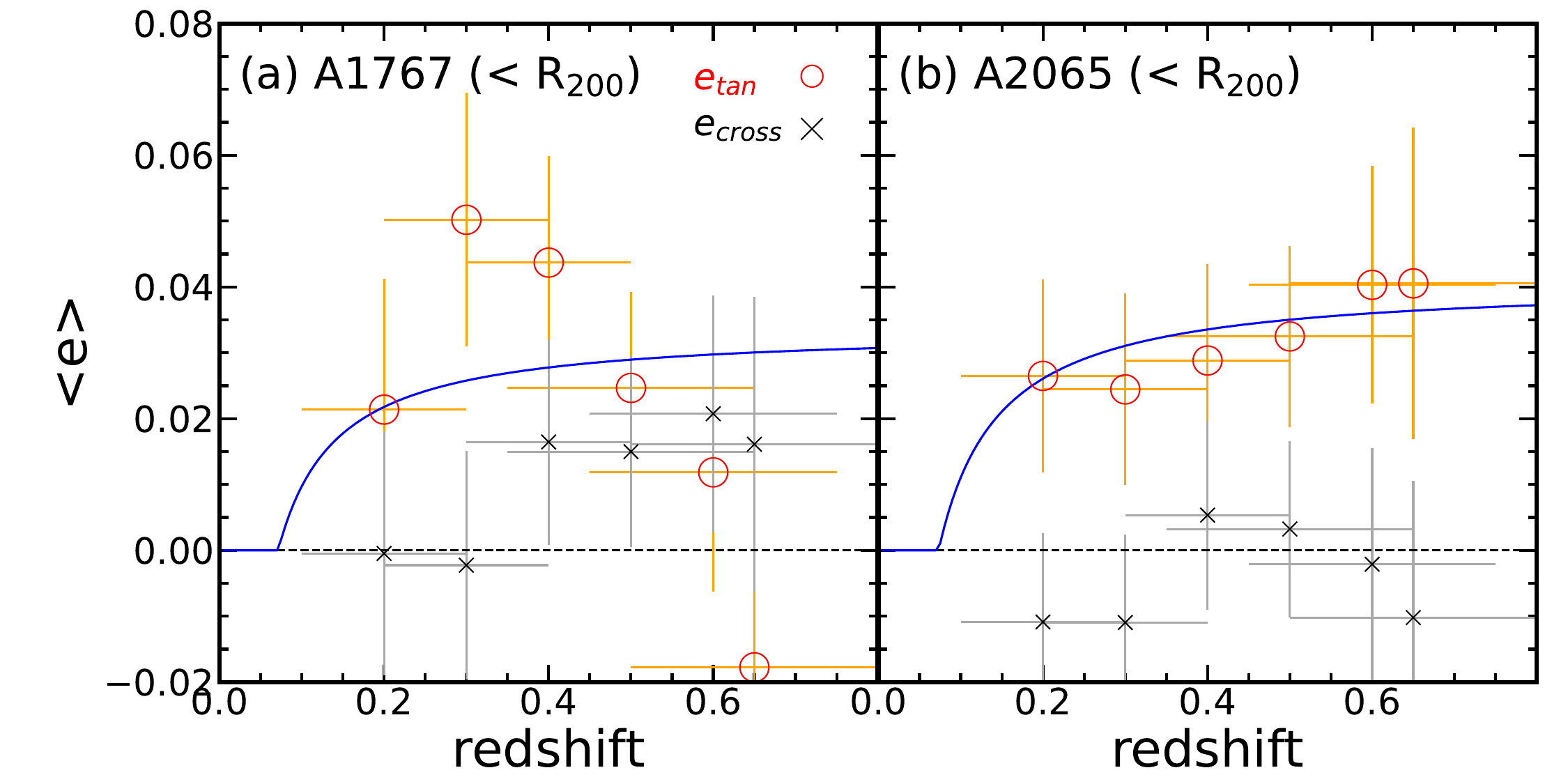}
\caption{Spectrotomographic signal as a function of redshift within $R_{200}$ of the projected distance from the centers of (a) A1767 and (b) A2065. Red and black symbols show the tangential and cross ellipticities, respectively. The error bar indicates the $1\sigma$ standard deviation divided by $\sqrt{N_{bin}}$, where $N_{bin}$ is the number of galaxies in each redshift bin. Blue curves display the predicted $\langle e_{tan} \rangle$ we derived based on the caustic mass of the two clusters.}
\label{fig:spectrotomography}
\end{figure*}

The blue curves in Figure \ref{fig:spectrotomography} show the predicted tomographic signals for the two clusters. We emphasize that these curves are {\bf not} fits to the ellipticity data; instead, they demonstrate consistency between the observed signal and the expectation based on independent measures of the cluster masses.

\subsection{Significance of the Tomographic Detections}

We use bootstrap resampling to compute the significance of the tomographic signal. We construct 100,000 realizations using 50\% of the galaxies in each realization. We sample six independent bins between $z = 0.1$ and $z = 0.8$. We then calculate the variance in the mean tomographic signal in each bin to compute the significance (adjusted for the sample size of the bootstrap). By construction, the signal in each bin is an independent measure of significance. We calculate the total significance by excluding the null hypothesis over the six independent bins.

For the samples including only galaxies with projected distance smaller than $R_{200}$ (Figure \ref{fig:spectrotomography}), the S/N for detection is $3.3\sigma$ for A1767 and $3.5\sigma$ for A2065. The significance of the tomographic signal is similar to the significance of the weak lensing map based on background objects with spectroscopy (Section \ref{sec:maps}). 

Because the redshift uncertainty is negligible compared to the width of our redshift bins, the assignment of galaxies to individual bins is effectively unambiguous. As a result, the uncertainty in the spectrotomographic signal is dominated by noise in the shape measurements within the shear bins. In the regime of interest, where the dispersion of ellipticity in each component is $\sim 30\%$ and the mean lensing-induced ellipticity is $\sim 4\%$, achieving a signal-to-noise ratio of $S/N \sim 3$ per bin requires approximately 500 galaxies within $R_{200}$ in each bin. While this is challenging, this number of spectroscopically confirmed background galaxies is attainable with multi-object fiber spectrographs on $8 - 10$ m class telescopes. Achieving substantially higher $S/N$ per bin would require stacking the signal over multiple clusters.

\subsection{Some Limitations of the Spectrotomography} \label{sec:limits}

\begin{figure*}
\centering
\includegraphics[scale=0.5]{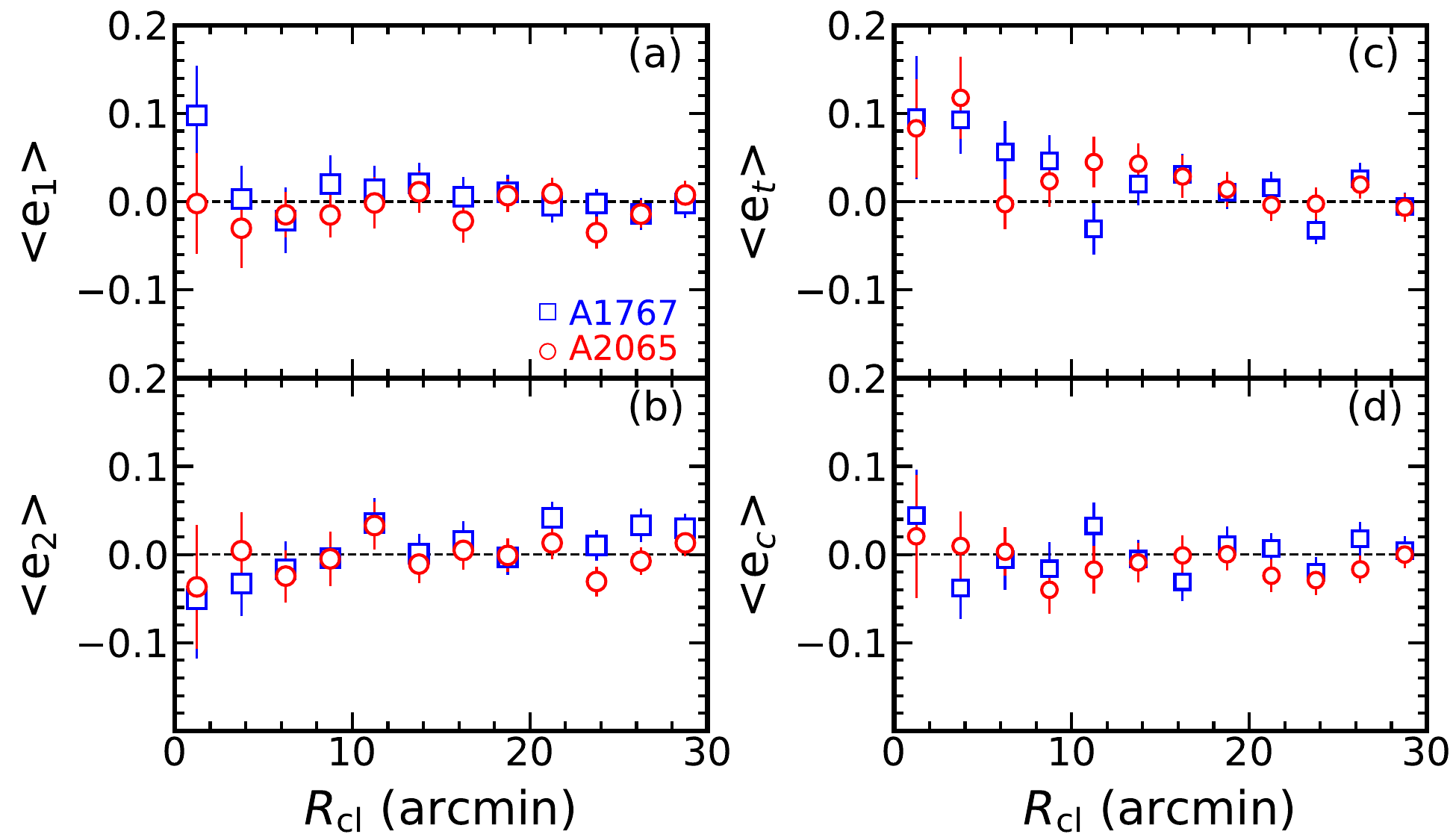}
\caption{Mean distributions of ellipticities ($e_{1}, e_{2}, e_{t}, e_{c}$) of spectroscopically identified background ($z > 0.1$) galaxies as a function of clustercentric distance. Blue squares and red circles display galaxies in the A1767 and A2065 field, respectively. }
\label{fig:spatial}
\end{figure*}

Although the S/N of the tomographic signals for the two clusters are similar, the tomographic trends in A1767 and A2065 differ significantly. For example, the A1767 tomographic signal is much stronger than the model prediction at $z < 0.4$ and much weaker at $z > 0.4$, while the A2065 tomographic signal is consistent with the model prediction across all redshift ranges. We thus examine the distributions of $e_{1}$ and $e_{2}$ to explore systematics that might account for these puzzling results. Figure \ref{fig:spatial} (a) and (b) show $\langle e_{1} \rangle$ and $\langle e_{2} \rangle$ as a function of clustercentric radius for both A1767 (blue) and A2065 (red). The value of $\langle e_{1} \rangle$ differs significantly from the zero expected value at small radii for A1767. In contrast, $\langle e_{2} \rangle$ is generally well-behaved.

Figure \ref{fig:spatial} (c) and (d) display $\langle e_{tan} \rangle$ and $\langle e_{cross} \rangle$ as a function of radius. The weak lensing signal is evident in the behavior of $\langle e_{tan} \rangle$ and $\langle e_{cross} \rangle$ for either cluster. However, the $\langle e_{tan} \rangle$ profile fluctuates at $R_{cl} > 5'$, which may propagate to the fluctuations in the tomographic signal in Figure \ref{fig:spectrotomography}. Given the small number of galaxies at each radius in each redshift bin, the assumption that galaxies in each redshift bin sample the same radial distribution is not well-supported, allowing radial biases in the tangential ellipticities to propagate into the tomographic signal. Unfortunately, there are not enough galaxies per bin in A1767 to test this hypothesis. However, if the small sample size is the cause, future larger samples of stacked clusters should not be sensitive to this bias. Similarly, an anisotropic cluster mass distribution could also produce variations in the tomographic signal for a single cluster. Together, these considerations suggest that cosmological constraints will require large stacked cluster samples.

Several potential sources may affect the radial distributions of the ellipticities. For example, contamination by light from extended bright member galaxies can affect accurate shape measurements for background galaxies near the cluster center. Analyses of the HSC/SSP data reduction have demonstrated that galaxy shapes, as inferred from second moments, are sensitive to details of the background light modeling in the outer isophotes, with this effect being most prominent near the cluster BCGs \citep{Aihara19}. However, there is no clean a priori method for rejecting contaminated background objects or for imposing a radial cut. The radial distributions of the ellipticities and lensing signals can also be affected by the cluster properties, including mass, dynamics, and substructure. Because the two clusters in our sample are generally well-relaxed and massive, our sample is too small to assess the impact of cluster properties on the tomographic and radial signals. We will discuss the effect of cluster properties with an extended cluster sample in Wright et al. (2026, in preparation).

\section{BUILDING A PLATFORM FOR SPECTROTOMOGRAPHY AND COSMOLOGY}\label{sec:future}

Spectrotomography of the three clusters, A1767, A2065 (this study), and A2029 \citep{dellAntonio20} previews a platform for spectrotomography based on 10 clusters with either DECam (2 clusters) or Subaru HSC photometry (8 clusters) along with extensive MMT Hectospec spectroscopy for all 10 systems in the redshift range 0.07 to 0.12 (Wright et al. in preparation). A1767 and A2065 bracket the mass range for the 10 cluster sample. The larger sample should enable a more robust understanding of the systematics that may affect the tomographic signal. In particular, a better understanding of the contamination of the background galaxy shape measurement by the extended light of central galaxies, and potential dependence of the cluster mass and dynamics. 

\cite{dellAntonio20} simulate future spectrotomographic observations with massively parallel spectrographs like Subaru/Prime Focus Spectrograph (PFS) \citep{Takada14}. They show that a PFS survey to a limiting $i = 22.7 $ (or $i = 23.7$) reaches a redshift of $z \sim 1.5$ for the sources. The predicted significance of the tomographic signal is $\sigma =7.4$ (and $\sigma = 11.7$ for a limiting $i = 22.7$). These results are comparable with current photo-$z$ methods. The uncertainty in the spectrotomographic mass estimate is currently large, but, in contrast with purely photometric weak lensing, no cluster members contaminate the signal. 

\cite{dellAntonio20} also emphasize that spectrotomography holds promise for determination of the cosmological parameters. Even a small sample of $\sim 50$ clusters provides a measurement of the angular diameter distance ratio as a function of background source redshift. The shape of this curve depends on the expansion history of the universe. Application of this purely geometric method for constraining the cosmological parameters is independent of other approaches \citep{Martinet15}.

\section{CONCLUSION}\label{sec:conc}

Weak lensing and redshift surveys are two of the most powerful tools of modern cosmology. We continue to build a platform for combining weak lensing with redshift surveys to obtain unbiased estimates of the masses of clusters of galaxies. We call this approach spectrotomography. \cite{dellAntonio20} report the first detection of the spectrotomographic signal for the rich cluster A2029.

Here we report the detection of the tomographic signal for two more clusters: A1767 and A2065. We base these detections on Subaru HSC imaging \citep{Aihara19} along with extensive redshift surveys of background galaxies with MMT Hectospec \citep{Sohn20}. The redshift surveys include 972 background galaxies for A1767 and 1563 for A2065. For comparisons, the first spectrotomographic detection for A2029 \citep{dellAntonio20} was based on 1517 background objects with spectroscopic redshifts.

We explore the $e_{tan}$ and $e_{cross}$ ellipticity distributions for the background galaxies in both systems and demonstrate that the lensing signal is present in the distributions of $e_{tan}$ relative to $e_{cross}$. The signal is absent for cluster members as expected.

We construct the first weak lensing maps based on galaxies with spectroscopic redshifts alone. Both A1767 and A2065 are detected in these weak lensing maps at a significance of $4.0 \sigma$ and $3.1 \sigma$ respectively. Although the significance is much less than the significance of detection in the full photometric weak lensing maps, the detections based on background galaxies with spectroscopic redshifts are a step toward the ultimate reduction of systematic errors due to cluster member contamination and background photometric redshift systematic errors with large samples of galaxy clusters.

We detect the tomographic signal for A1767 ($M_{200} = 5.97 \times 10^{14}$ M$_{\odot}$) and A2065 ($M_{200} = 10.19 \times 10^{14}$ M$_{\odot}$) at a level of $3.1\sigma$ and $3.5\sigma$ within $R_{200}$, respectively. The significance of our first spectrotomographic detection, A2029 ($M_{200} = 8.45 \times 10^{14}$ M$_{\odot}$) is 3.9$\sigma$ based on DECam photometry and a comparable number of background redshifts. These results demonstrate both the potential of spectrotomography and the care required to interpret the results. 

Issues that may affect the tomographic signal include (1) accurate measurement of galaxy shapes under the influence of extended massive cluster galaxies (e.g., Brightest Cluster Galaxies), (2) the limited number of spectroscopic background galaxies at higher redshifts, (3) cluster properties including their mass and dynamical status. In particular, the large fluctuations of both $\langle e_{tan} \rangle$ and $\langle e_{cross} \rangle$ observed in A1767 are absent in A2029 and A2065. Expanding the sample to include systems spanning a wider range of background complexity and spectroscopic sampling will allow a more systematic assessment of these effects.

The great advantage of spectrotomography relative to tomography based on photometric redshifts lies in the precise determination of angular diameter distance ratios. In photometric redshift-based tomographic measurements, the presence of unidentified redshift outliers (``catastrophic errors") creates a smoothing effect on the shear-redshift plot that can bias the cosmological parameter estimation. In contrast, spectroscopically determined redshifts have essentially no bias. 

Massively multiplexed fiber spectrographs including DESI \citep{DESI_DR1}, Subaru/PFS \citep{Takada14}, and DESI II \citep{Schlegel22} promise a huge advance in spectrotomography. These fiber-fed systems often have fiber exclusion zones that make complete redshift surveys difficult to achieve, but the tomographic signal depends simply on the number of spectra available behind clusters. Incomplete sampling of the lensing distortion field behind clusters and its asymmetry in any one cluster (due to infalling groups and other substructures in the cluster vicinity) may, however, alter the angular diameter distance ratio diagram.  Measuring the signal over a sample of clusters at each redshift (and potentially cluster mass) averages over significant substructures to reduce bias in the tomographic measurement. Upcoming deep surveys including LSST \citep{Ivezic19}, the EUCLID wide survey \citep{Euclid}, and the Roman Space Telescope High Latitude Wide-Area Survey \citep{Wang22} will map out tens of thousands of galaxy clusters, providing abundant targets for spectrotomographic measurement. In the coming years, it will be possible to obtain hundreds of redshifts for galaxies behind the virial region of clusters out to $z\sim 0.6$, and many thousands behind nearby ($z\sim 0.1-0.2$) clusters. 


\acknowledgments

We thank the anonymous referee who provided thoughtful comments that significantly improve the quality of the manuscript. We also thank Yousuke Utsumi for his expert advice on all aspects of weak lensing analysis. We also thank Lily Wright and Ivana Damjanov for insightful comments. This work was supported by Creative-Pioneering Researchers Program through Seoul National University. This work was also supported by the Global-LAMP Program of the National Research Foundation of Korea (NRF) grant funded by the Ministry of Education (No. RS-2023-00301976). 

This paper is based on data from the Hyper Suprime-Cam Legacy Archive (HSCLA), which is operated by the Subaru Telescope. The original data in HSCLA was collected at the Subaru Telescope and retrieved from the HSC data archive system, which is operated by the Subaru Telescope and Astronomy Data Center at National Astronomical Observatory of Japan. The Subaru Telescope is honored and grateful for the opportunity of observing the Universe from Maunakea, which has the cultural, historical and natural significance in Hawaii.

This paper makes use of software developed for the Vera C. Rubin Observatory. We thank the observatory for making their code available as free software at http://dm.lsst.org.

The Pan-STARRS1 Surveys (PS1) and the PS1 public science archive have been made possible through contributions by the Institute for Astronomy, the University of Hawaii, the Pan-STARRS Project Office, the Max Planck Society and its participating institutes, the Max Planck Institute for Astronomy, Heidelberg, and the Max Planck Institute for Extraterrestrial Physics, Garching, The Johns Hopkins University, Durham University, the University of Edinburgh, the Queen’s University Belfast, the Harvard-Smithsonian Center for Astrophysics, the Las Cumbres Observatory Global Telescope Network Incorporated, the National Central University of Taiwan, the Space Telescope Science Institute, the National Aeronautics and Space Administration under grant No. NNX08AR22G issued through the Planetary Science Division of the NASA Science Mission Directorate, the National Science Foundation grant No. AST-1238877, the University of Maryland, Eotvos Lorand University (ELTE), the Los Alamos National Laboratory, and the Gordon and Betty Moore Foundation.

Funding for SDSS-III has been provided by the Alfred P. Sloan Foundation, the Participating Institutions, the National Science Foundation, and the U.S. Department of Energy Office of Science. The SDSS-III website is http://www.sdss3.org/. SDSS-III is managed by the Astrophysical Research Consortium for the Participating Institutions of the SDSS-III Collaboration, including the University of Arizona, the Brazilian Participation Group, Brookhaven National Laboratory, University of Cambridge, Carnegie Mellon University, University of Florida, the French Participation Group, the German Participation Group, Harvard University, the Instituto de Astro sica de Canarias, the Michigan State/Notre Dame/JINA Participation Group, Johns Hopkins University, Lawrence Berkeley National Laboratory, Max Planck Institute for Astrophysics, Max Planck Institute for Extraterrestrial Physics, New Mexico State University, New York University, The Ohio State University, The Pennsylvania State University, University of Portsmouth, Princeton University, the Spanish Participation Group, University of Tokyo, University of Utah, Vanderbilt University, University of Virginia, University of Washington, and Yale University.


\bibliography{ms}


\end{document}